# Enhancing Decentralization in Blockchain Decision-Making Through Quadratic Voting and Its Generalization

Lyudmila Kovalchuk[1,2], Mariia Rodinko[1,3], Roman Oliynykov[1,3], Andrii Nastenko[1,4], Dmytro Kaidalov[1], and Kenric Nelson[5]

[1]IOG, Singapore
[2]G. E. Pukhov Institute for Modelling in Energy Engineering, Ukraine
[3]V. N. Karazin Kharkiv National University, Ukraine
[4]Kharkiv National University of Radio Electronics, Ukraine
[5]Photrek, USA

**Abstract.** This study explores the application of Quadratic Voting (QV) and its generalization to improve decentralization and effectiveness in blockchain governance systems. The conducted research identified three main types of quadratic (square root) voting. Two of them pertain to voting with a split stake, and one involves voting without splitting. In split stakes, Type 1 QV applies the square root to the total stake before distributing it among preferences, while Type 2 QV distributes the stake first and then applies the square root. In unsplit stakes (Type 3 QV), the square root of the total stake is allocated entirely to each preference.

The presented formal proofs confirm that Types 2 and 3 QV, along with generalized models, enhance decentralization as measured by the Gini and Nakamoto coefficients. A pivotal discovery is the existence of a "threshold" stakeholder whose relative voting ratio increases under QV compared to linear voting, while smaller stakeholders also gain influence. The generalized QV model allows flexible adjustment of this threshold, enabling tailored decentralization levels.

Maintaining fairness, QV ensures that stakeholders with higher stakes retain a proportionally greater voting ratio while redistributing influence to prevent excessive concentration. It is shown that to preserve fairness and robustness, QV must be implemented alongside privacy-preserving cryptographic voting protocols, as voters casting their ballots last could otherwise manipulate outcomes.

The generalized QV model, proposed in this paper, enables algorithmic parametrization to achieve desired levels of decentralization for specific use cases. This flexibility makes it applicable across diverse domains, including user interaction with cryptocurrency platforms, facilitating community events and educational initiatives, and supporting charitable activities through decentralized decision-making.

**Introduction**

The literature surrounding blockchains has expanded significantly, particularly in terms of the mechanisms that facilitate decentralization. One of the primary discussions revolves around the economic incentives that govern validator participation. Research by Bonneau et al. (2015) emphasizes how economic models can either promote or inhibit decentralization, particularly in Proof-of-Stake (PoS) systems where validators are rewarded based on their stake. Authors suggest that designing incentive structures that encourage a broader distribution of stakes can mitigate centralization tendencies. The role of governance in PoS blockchains has been highlighted by various authors, including Buterin (2014), who argues that decentralized governance mechanisms are essential for maintaining network integrity. Buterin expands on this by proposing specific governance models that integrate community input and decision-making processes, thereby fostering a more inclusive environment.

Kiayias and Lazos (2022) say that in PoS systems, voting weight is often tied to stake or wealth, leading to issues such as enthusiastic or high-contributing participants having less influence. Similarly, in Proof-of-Work (PoW), hashing power doesn't always reflect contributions. Analyzing decentralization can help understand these power imbalances. Juodis et al. (2024) concluded that Ethereum's transition to the PoS consensus protocol had minimal effect on wealth decentralization. In contrast, Layer 2 blockchains such as Arbitrum, Optimism, and Polygon demonstrate markedly higher centralization across all analyzed metrics. Ovezik et al. (2025) highlight the significance of decentralization while emphasizing the absence of a universally accepted method for its measurement. Their findings question the long-standing belief within the blockchain community that increased participation leads to greater decentralization.

The complexities of decision-making in collective environments have prompted the exploration of innovative voting mechanisms that better capture the preferences of participants. One such mechanism is Quadratic Voting, a novel approach that seeks to address the limitations of traditional voting systems. Traditional methods often fail to represent the intensity of voters' preferences, leading to outcomes that may not reflect the true desires of the electorate. Quadratic Voting (QV), introduced by Posner and Weyl (2014), offers a solution by allowing individuals to express the strength of their preferences through a system of "votes" that can be purchased at a cost that increases quadratically with the number of votes cast.

In blockchain governance, QV helps address the issue of dominant stakeholders controlling the majority of votes. By allowing participants to express the intensity of their preferences, QV ensures that decisions reflect both the number and the strength of supporters' preferences (Dimitri 2022). QV involves two scenarios: split and unsplit stakes. In a split stake scenario, voters distribute their voting credits among multiple options based on the intensity of their preferences. This approach allows voters to allocate more votes to favored options while still supporting others to a lesser extent. In contrast, an unsplit stake scenario requires voters to allocate all their voting credits to each preferable option. For Quadratic Voting with split stakes, there are also two scenarios. In the first scenario (Type 1 QV), the square root of the stake is calculated first, and this value is then distributed among various preferences. In the second scenario (Type 2 QV), the stake is distributed first, and the square root of the resulting values is taken. In the unsplit stake scenario (Type 3 QV), the square root of the entire stake is calculated, and this value is fully allocated to each preference.

Numerous papers have examined the properties of various types of QV. Let's review some of the most notable results in this area. Dimitri (2022) considers QV in PoS-based blockchains. It focuses on a game theoretical approach to determine the optimal allocation of votes in rounds (the stake of each voter is distributed among several rounds of voting). It seems that we may consider preferences/proposals instead of rounds. The possibility of Sybil attacks is also investigated by Dimitri (2022). The paper contains a good description of the model and proposes several statements about Nash equilibrium for some special cases. Additionally, it introduces an alternative approach to QV, named AQV (Type 2 QV): the sum of "voting stake shares" for the voter is equal to the square root of his total stake.

Robey (2022) defines Type 1 QV and explains its advantage: it favors minorities over large stakeholders. However, it does not present any new results or mathematical analysis. Bobby (2022) defines Type 3 QV and emphasizes that QV is susceptible to Sybil attacks, highlighting the need for identification mechanisms to prevent such vulnerabilities. Miller et al. (2022) consider Quadratic Funding (QF) and QV. The approach is similar to Type 1 QV. The paper primarily discusses QF, with a brief mention of QV. It highlights that QV is resistant to collusion, as it requires participants to pay quadratically for buying votes.

Fritsch et al. (2022) investigate the measurement of decentralization using the Gini coefficient and Nakamoto coefficient, providing a comparison of these values for Compound, Uniswap, and ENS. Langer et al. (2010) discuss the concepts of anonymity and verifiability in voting by examining (un)linkability. It clarifies that both can be described in terms of linkability: anonymity requires unlinkability between the voter and their vote, whereas verifiability requires linkability between voters and the election result.

Lalley and Weyl (2014, 2018) explore game-theoretical and probability-theoretical approaches in voting, with an extensive analysis of QV. It demonstrates that QV is as efficient, if not more efficient, than traditional voting systems. Efficiency refers to the ability of a community to make decisions that improves the overall welfare. Authors consider QV as voting based on a quadratic pricing rule. They show that QV is an optimal intermediate point between the extremes of dictatorship and majority rule. They prove that such power (power 2) in price for a vote preference optimizes the price-taking equilibrium for each person's and the community's utility: "a vote pricing rule is robustly optimal if and only if it is quadratic".

Wallis (2014) provides an overview of several key types of voting systems, making it a valuable resource for those starting to work in the field of e-voting. Akinbohun et al. (2023) conduct a literature review of blockchain-based voting systems, focusing primarily on the criteria for selecting articles and comparing them based on formal indicators such as term occurrence percentages. Vasiljev

(2014) argues that cardinal voting, which uses valued scales, is superior to ordinal voting, where preferences are ranked in descending order. It includes numerous useful definitions related to cardinal voting and presents straightforward mathematical explanations.

Posner and Weyl (2015) state the merits of QV: it solves the tyranny-of-the-majority problem, the major defect of majority rule, and it does so without creating gridlock. Chandar and Weyl (2019) compare QV and 1 person – 1 vote voting system. The authors show that 1 person – 1 vote may be better than QV in some special scenarios. Also, the results obtained allow authors to suggest that in highly unequal societies, 1 person – 1 vote or QV with artificial currency may give superior efficiency to QV with real currency.

Benjamin et al. (2017) explore the relationship between the normalized gradient addition mechanism and QV. Both are social choice mechanisms that impose quadratic budget constraints on individuals, but they are used in different contexts. The paper also offers a formal analysis of QV when votes are paid for with abstract tokens distributed equally by the mechanism designer, instead of money.

Kho et al. (2022) performed a comprehensive comparison analysis between the various e-voting approaches, namely, mix-net-based e-voting, homomorphic e-voting, blind signature-based e-voting, blockchain-based e-voting, post-quantum e-voting, and hybrid e-voting. The development of the respective approaches was reviewed, and a detailed comparison was conducted on the specific schemes in each approach. Also, some practical considerations are discussed and some potential research directions are underlined. Strandberg et al. (2025) present a short simple text about the fairness problem in voting described with Arrow's Impossibility Theorem: it is impossible to fulfill all of the three above features (Unanimity, Non-dictatorship, IIA) at the same time in any ranked voting system.

Hajian Berenjestanaki et al. (2023) provide a comprehensive review of blockchain-based e-voting systems, analyzing 252 selected papers. It highlights that security, transparency, and decentralization are the main benefits, while privacy, verifiability, efficiency, trustworthiness, and auditability also receive attention but are not the primary focus. The paper also notes a lack of emphasis on accessibility, compatibility, availability, and usability.

Tamai and Kasahara (2024) investigate the whale problem in Decentralized Autonomous Organizations (DAOs). They analyze QV, designed to prevent voting power concentration, and show it is less resistant to collusion compared to Linear Voting. To address this, they propose a mechanism combining QV with vote escrow tokens, mitigating the whale problem and enhancing resilience to collusion. Lei et al. (2024) incorporate QV into Delegated PoS (DPoS) to improve voting power distribution, encourage participation from smaller stakeholders, and reduce influence concentration. Robustness is enhanced through admission rules and vote similarity detection to counter Sybil Attacks. Simulations show QDPoS boosts voter turnout, reduces stake centralization, and strengthens decentralization. Game theory analysis confirms its ability to diversify voting preferences, fostering a balanced and resilient consensus for Web 3.0.

Despite many papers investigating QV, there are enough tasks in this area that require solutions. This paper aims to explore the theoretical foundations and practical implications of quadratic voting as a means to enhance democratic participation and decision-making processes.

**Section 1** defines the basic concepts and introduces the types of QV. **Section 2** explores the properties of relative voting ratio for Types 2 and 3 QV, along with the changes in RVR when transitioning from Linear Voting to Quadratic Voting, and examines the generalized QV model aimed at enhancing decentralization. **Section 3** presents formal proofs for both the Gini coefficient and the Nakamoto coefficient, demonstrating that Type 2 QV fosters greater voter equality compared to the 1 coin – 1 vote scheme and outlines the conditions under which this holds true. Additionally, the mechanism detects participants who may attempt to exploit the system through collusion-based attacks in Type 1 QV. **Section 4** analyzes Types 1 and 2 QV, emphasizing the necessity of deploying QV and its generalized form with a privacy-preserving cryptographic voting protocol to prevent strategic manipulation by voters casting their votes after observing prior results. **Section 5** assesses the losses and profits of QV for major stakeholders and their influence on voting outcomes. **Section 6** introduces an algorithm to achieve a desired level of maximal stake impact. The final section summarizes the findings and draws conclusions.

## 1. Basic definitions

Define $V = \{V^{(1)}, \ldots, V^{(n)}\}$ the set of Voters, and $s^{(1)}, \ldots, s^{(n)}$ their corresponding voting stakes, $s^{(1)} + \ldots + s^{(n)} = s$, where $s$ is their total voting stake. Sometimes (with additional notes) we will use normalized voting stakes for simplicity. Note that Voters or other Stakeholders may also have additional stake which doesn't take part in voting.

Let $p_1, \ldots, p_m$ be alternatives (we will call them *proposals*) which take part in the current voting round (or *fund* in treasury systems).

Let $f, g: \mathbf{R}_{\geq 0} \to \mathbf{R}_{\geq 0}$ be some increasing functions.

Define $B = \{B^{(1)}, \ldots, B^{(n)}\}$ the set of ballots that Voters fill, where
$$B^{(i)} = (b_1^{(i)}, \ldots, b_m^{(i)}), \quad i = \overline{1, n},$$
is profile of Voter $V^{(i)}$ – vector which describes the results of his voting.

Note that a superscript also refers to the voter's index and a subscript refers to the proposal's number.

We will call a voting scheme (VS) *the yes-no-abstain voting scheme with unsplit stake*, if $b_1^{(i)}, \ldots, b_m^{(i)} \in \{-g(s^{(i)}), 0, g(s^{(i)})\}$, and *the yes-abstain voting scheme with unsplit stake*, if $b_1^{(i)}, \ldots, b_m^{(i)} \in \{0, g(s^{(i)})\}$.

If the values $b_1^{(i)}, \ldots, b_m^{(i)}$ have to satisfy the requirement: $\sum_{l=1}^{m} |b_l^{(i)}| = g(s^{(i)})$ (from where we can conclude that $b_1^{(i)}, \ldots, b_m^{(i)} \in [-g(s^{(i)}), g(s^{(i)})]$), then we call this VS *the VS with split stake*. If also $b_1^{(i)}, \ldots, b_m^{(i)} \in [-g(s^{(i)}), g(s^{(i)})]$, we call it *the yes-no-abstain voting scheme with split stake*; if $b_1^{(i)}, \ldots, b_m^{(i)} \in [0, g(s^{(i)})]$, we call it *the yes-abstain voting scheme with split stake*.

Now $g(s^{(i)})$ represents the real value that $V^{(i)}$ may split among the preferable proposals, and we will call this value *the voting credit with respect to function $g$* and denote it $vc_g^{(i)}$.

In practice, we may also set some additional restrictions on these values, connected with their discretization (for example, they may take integer values, expressed in the number of coins).

For each proposal $p_i$, define its *score*, $score(p_i)$, $i = \overline{1, m}$, as
$$score(p_i) = \sum_{j=1}^{n} b_i^{(j)},$$
and its *voting score w.r.t. function $f$* as
$$vscore(p_i) = \sum_{j=1}^{n} sign(b_i^{(j)}) f(|b_i^{(j)}|).$$

Two of the most widely used examples of functions $f$ and $g$ are:
$$f, g(x) = x \quad \text{and} \quad f, g(x) = \sqrt{x}.$$

We call a voting scheme *the linear VS*, if $g(x) = x$ and $f(x) = x$.

For example, in these terms, the current Catalyst VS (Cardano treasury system) is **the yes-abstain linear VS with unsplit stake**.

The graphical representation of the voting model is presented in Fig. 1.

**Definition 1.** Using our designations, we can define the three main types of QV, which we will investigate below:

**type 1 of QV:** QV with split stake, where $g(x) = x$, $f(x) = \sqrt{x}$;
**type 2 of QV:** QV with split stake, where $g(x) = \sqrt{x}$, $f(x) = x$;
**type 3 of QV:** QV with unsplit stake, where $g(x) = \sqrt{x}$, $f(x) = x$.

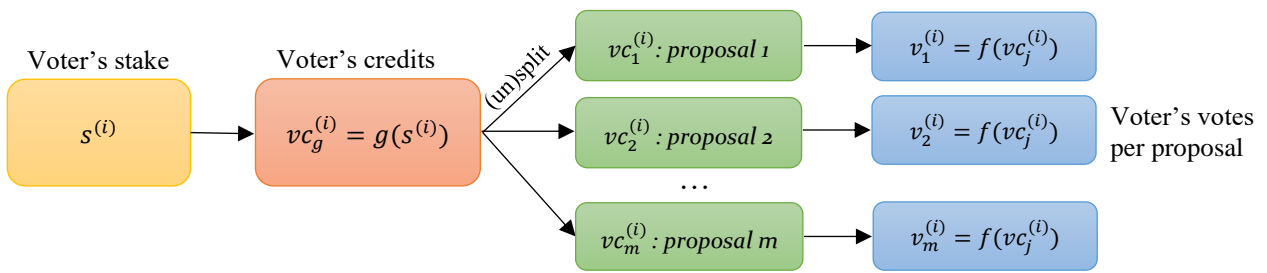

Fig. 1 – The basic voting model

Note that though we define QV type 3 (with unsplit stake) based on the definition of QV type 2, it may be obtained either from type 1 or type 2 (with split stake) with only change connected with the type of stake usage.

## 2. Properties of relative voting ratio in QV Type 2 and Type 3

In this section we analyze the properties of the relative voting ratio (RVR) for QV types 2 and 3, and how the RVR changes when moving from Linear Voting to Quadratic Voting. The main difference between these two voting schemes, type 2 and type 3, is that in type 3 each voter may use their stake many times, voting for as many proposals as they want without stake decreasing. This leads to the fact that the RVR of a voter depends not only on his stake value but also on how many proposals he supports. Thus, in QV type 3 the voter with a smaller stake, who votes for a lot of proposals, may have a larger RVR than the voter with a larger stake, who supports only a few proposals.

The RVR influences the power of voters nonlinearly. The power of a voter is measured by the probability of the voter to change the outcome of the community's decision. The Banzhaf Power Index and the Shapley-Schubik Power Index are the two most common measures. Each of these examines the power set of possible coalitions between voters. If a voter's presence in a coalition changes that coalition from a losing to a winning coalition that voter is a swing voter. The probability of being a swing voter is that voter's power. Each index makes subtle distinctions in defining the set of swing outcomes. While our analysis is restricted to the RVR, it's important to understand this is not directly the power. In fact, Penrose's discovery that the power grows approximately quadratically with increases in votes, is part of the underlying significance of the quadratic voting method.

### 2.1. Properties of relative voting ratio in QV Type 2

Here we describe some properties of QV-2, which show an increase of some decentralization characteristics in comparison with linear VS.

Define *relative voting ratio* (RVR) of voter $V^{(i)}$ as $r_g^{(i)} = \frac{g(s^{(i)})}{\sum_{j=1}^n g(s^{(j)})}$. In particular case, when $g(x) = \sqrt{x}$ (as in QV-2) we will write $r_{QV}^{(i)} = \frac{\sqrt{s^{(i)}}}{\sum_{j=1}^n \sqrt{s^{(j)}}}$, and for linear VS we will write $r_{LV}^{(i)} = \frac{s^{(i)}}{\sum_{j=1}^n s^{(j)}} = \frac{s^{(i)}}{s}$.

Note that both types of QV, type 1 and type 2, are vulnerable to Sybil attacks, but QV-3 is not.

Here we are going to show that in QV-2 the RVR decreases for "whales" and increases for voters with small stake, keeping the order of initial voting ratio.

Define $\eta^{(i)} = \frac{r_{QV}^{(i)}}{r_{LV}^{(i)}}$ – the value which characterizes the relation of RVR in QV w.r.t. RVR in in Linear Voting for $i$-th voter. If $\eta^{(i)} > 1$, then the RVR increases when moving to QV, and vice versa.

In our definitions, we have the next Proposition.

**Proposition 1 (property of RVR in QV-2).**
In our designations, let $s^{(1)} < \ldots < s^{(n)}$. Then the next statements are true.
1. For $1 \leq i < j \leq n$, and for each increasing function $g$, $r_g^{(i)} < r_g^{(j)}$.
2. In our designations, $r_{QV}^{(1)} > r_{LV}^{(1)}$ and $r_{QV}^{(n)} < r_{LV}^{(n)}$.
3. If for some $1 \leq k \leq n$, $r_{QV}^{(k)} > r_{LV}^{(k)}$, then for each $1 \leq l < k$: $r_{QV}^{(l)} > r_{LV}^{(l)}$.
4. If for some $1 \leq k \leq n$, $r_{QV}^{(k)} < r_{LV}^{(k)}$, then for each $k < l \leq n$: $r_{QV}^{(l)} < r_{LV}^{(l)}$.
5. If for some $1 \leq i < k \leq n$ we have $s^{(i)} < s^{(k)}$, then $\eta^{(i)} > \eta^{(k)}$.
6. For $1 \leq i \leq n$, the relation $\eta^{(i)} > 1$ holds iff $\sqrt{s^{(i)}} < \frac{\sum_{j=1}^n s^{(j)}}{\sum_{j=1}^n \sqrt{s^{(j)}}}$.

*Informally speaking, the 1st statement says that VS (for any increasing g) preserves the increasing order of stake. The 2nd says that for the voter with the largest stake, his RVR always decreases in QV-2, and for the voter with the smallest stake, his RVR always increases in QV-2. The 3rd statement says that if for some voter his RVR in QV-2 is larger than for linear VS, then the same is true for all voters with smaller stake. Analogically, the 4th statement says that if for some voter his RVR in QV-2 is smaller than for linear VS, then the same is true for all voters with larger stake. The 5th statement shows that the smaller the stake of the voter, the more significantly increases his RVR when moving to QV. The 6th statement gives the criterion of increasing RVR.*

*Proof.*

1. According to our designations, $s^{(i)} < s^{(j)}$, so $g(s^{(i)}) < g(s^{(j)})$ and $r_g^{(i)} = \frac{g(s^{(i)})}{\sum_{j=1}^n g(s^{(j)})} < \frac{g(s^{(j)})}{\sum_{j=1}^n g(s^{(j)})} = r_g^{(j)}$.

2. According to our designations, $s^{(1)} < s^{(j)}$ for each $2 \leq j \leq n$. Then

$$r_{QV}^{(1)} = \frac{\sqrt{s^{(1)}}}{\sum_{j=1}^n \sqrt{s^{(j)}}} = \frac{1}{\sum_{j=1}^n \sqrt{\frac{s^{(j)}}{s^{(1)}}}} > \frac{1}{\sum_{j=1}^n \frac{s^{(j)}}{s^{(1)}}} = \frac{s^{(1)}}{\sum_{j=1}^n s^{(j)}} = r_{LV}^{(1)},$$

using that $\frac{s^{(j)}}{s^{(1)}} > 1$ and then $\frac{s^{(j)}}{s^{(1)}} > \sqrt{\frac{s^{(j)}}{s^{(1)}}}$.

Analogically, according to our designations, $s^{(n)} > s^{(j)}$ for each $1 \leq j \leq n-1$. Then

$$r_{QV}^{(n)} = \frac{\sqrt{s^{(n)}}}{\sum_{j=1}^n \sqrt{s^{(j)}}} = \frac{1}{\sum_{j=1}^n \sqrt{\frac{s^{(j)}}{s^{(n)}}}} < \frac{1}{\sum_{j=1}^n \frac{s^{(j)}}{s^{(n)}}} = \frac{s^{(n)}}{\sum_{j=1}^n s^{(j)}} = r_{LV}^{(n)},$$

using that $\frac{s^{(j)}}{s^{(n)}} < 1$ and then $\frac{s^{(j)}}{s^{(n)}} < \sqrt{\frac{s^{(j)}}{s^{(n)}}}$.

3. Let for some $1 \leq k \leq n$, $r_{QV}^{(k)} > r_{LV}^{(k)}$, which means that

$$r_{QV}^{(k)} = \frac{\sqrt{s^{(k)}}}{\sum_{j=1}^n \sqrt{s^{(j)}}} > r_{LV}^{(k)} = \frac{s^{(k)}}{\sum_{j=1}^n s^{(j)}}.$$

Then for each $1 \leq l < k$ we have:

$$r_{QV}^{(l)} = \frac{\sqrt{s^{(l)}}}{\sum_{j=1}^n \sqrt{s^{(j)}}} = \frac{\sqrt{s^{(l)}}}{\sqrt{s^{(k)}}} \cdot \frac{\sqrt{s^{(k)}}}{\sum_{j=1}^n \sqrt{s^{(j)}}} > \sqrt{\frac{s^{(l)}}{s^{(k)}}} \cdot r_{LV}^{(k)} >$$

$$> \frac{s^{(l)}}{s^{(k)}} \cdot \frac{s^{(k)}}{\sum_{j=1}^n s^{(j)}} = \frac{s^{(l)}}{\sum_{j=1}^n s^{(j)}} = r_{LV}^{(l)},$$

using $\sqrt{\frac{s^{(l)}}{s^{(k)}}} > \frac{s^{(l)}}{s^{(k)}}$.

4. Let for some $1 \leq k \leq n$, $r_{QV}^{(k)} < r_{LV}^{(k)}$, which means that

$$r_{QV}^{(k)} = \frac{\sqrt{s^{(k)}}}{\sum_{j=1}^n \sqrt{s^{(j)}}} < r_{LV}^{(k)} = \frac{s^{(k)}}{\sum_{j=1}^n s^{(j)}}.$$

Then for each $k < l \leq n$ we have:

$$r_{QV}^{(l)} = \frac{\sqrt{s^{(l)}}}{\sum_{j=1}^n \sqrt{s^{(j)}}} = \frac{\sqrt{s^{(l)}}}{\sqrt{s^{(k)}}} \cdot \frac{\sqrt{s^{(k)}}}{\sum_{j=1}^n \sqrt{s^{(j)}}} < \sqrt{\frac{s^{(l)}}{s^{(k)}}} \cdot r_{LV}^{(k)} <$$

$$< \frac{s^{(l)}}{s^{(k)}} \cdot \frac{s^{(k)}}{\sum_{j=1}^n s^{(j)}} = \frac{s^{(l)}}{\sum_{j=1}^n s^{(j)}} = r_{LV}^{(l)},$$

using $\sqrt{\frac{s^{(l)}}{s^{(k)}}} < \frac{s^{(l)}}{s^{(k)}}$.

5. Let for some $1 \leq i < k \leq n$ we have $s^{(i)} < s^{(k)}$. Then

$$\eta^{(i)} = \frac{r_{QV}^{(i)}}{r_{LV}^{(i)}} = \frac{\sqrt{s^{(i)}}}{\sum_{j=1}^{n}\sqrt{s^{(j)}}} \cdot \frac{\sum_{j=1}^{n} s^{(j)}}{s^{(i)}} = \frac{\sum_{j=1}^{n} s^{(j)}}{\sum_{j=1}^{n}\sqrt{s^{(j)}}} \cdot \frac{1}{\sqrt{s^{(i)}}} >$$

$$> \frac{\sum_{j=1}^{n} s^{(j)}}{\sum_{j=1}^{n}\sqrt{s^{(j)}}} \cdot \frac{1}{\sqrt{s^{(k)}}} = \frac{r_{QV}^{(k)}}{r_{LV}^{(k)}} = \eta^{(k)}.$$

The property 6 may be proved by direct calculations.
Proposition is proved.  □

**Generalization:** Proposition 1 is true for arbitrary increasing function $g: R_{\geq 0} \to R_{\geq 0}$ with the next property: $g(x) \geq x$ for $x \in [0,1]$ and $g(x) \leq x$ for $x \geq 1$. In particular, it is true for the function $g(x) = x^\gamma$ for $\gamma \in (0,1)$.

**Definition 2.** We will call the voting scheme with $g(x) = x^\gamma$ for $\gamma \in (0,1)$ as the gamma-power voting scheme, or gamma-power voting (GPV). QV is a particular case of GPV with $\gamma = 0.5$.

### 2.2. Properties of relative voting ratio in QV Type 3

One of the alternative approaches to QV may be QV with an unsplit stake. Note that in this case QV-1 and QV-2 are the same because the stake is not split. We will call this QV-3. In this VS, each voter votes with his total stake for an arbitrary number of proposals. So, his relative voting ratio depends not only on his stake but also on the number of proposals, for which each voter votes.

The advantages of this approach are the next:
1) simplicity in usage: the voter need not decide how to split his stake between proposals;
2) simplicity in calculation;
3) (relative) simplicity in providing privacy-preserving properties (VS with split stake needs much more zk-algorithms and protocols to achieve the same level of privacy-preserving).

The possible disadvantages may be the next:
1) bribery: as a voter doesn't lose any part of the stake, voting for proposals, he may be bribed by an arbitrary number of adversaries;
2) the voter who has a significant minority of stake may vote for a lot of proposals, which possess the necessary threshold.

Here we describe some properties of QV-3, which show an increase of some decentralization characteristics in comparison with linear VS.

Let in some voting process each voter $V^{(i)}$ votes (with his unsplit stake) for $c_i$ proposals. Define *relative voting ratio* (RVR) of voter $V^{(i)}$ as $u_g^{(i)} = \frac{c_i \cdot g(s^{(i)})}{\sum_{j=1}^{n} c_j \cdot g(s^{(j)})}$. (Here we use designation $u_g^{(i)}$ to emphasize that this VS is with UNSPLIT stake.) In particular case, when $g(x) = \sqrt{x}$ (as in QV-2 or QV-3) we will write $u_{QV}^{(i)} = \frac{c_i \cdot \sqrt{s^{(i)}}}{\sum_{j=1}^{n} c_j \cdot \sqrt{s^{(j)}}}$, and for linear VS we will write $u_{LV}^{(i)} = \frac{c_i \cdot s^{(i)}}{\sum_{j=1}^{n} c_j \cdot s^{(j)}}$.

Analogously to QV-2, define for QV-3 the value $\eta^{(i)} = \frac{u_{QV}^{(i)}}{u_{LV}^{(i)}}$, which characterizes the relation of RVR in QV w.r.t. RVR in Linear Voting for $i$-th voter. If $\eta^{(i)} > 1$, then the RVR increases when moving to QV, and vice versa.

Here we are going to formulate and prove the series of statements, which are analogical to some statements from Proposition 1. In particular, we show that in QV-3 the RVR decreases for "whales" and increases for voters with small stakes, keeping the order of initial voting ratio.

**Proposition 2 (properties of RVR in QV-3).**
In our designations, let $s^{(1)} < \ldots < s^{(n)}$. Then the next statements are true.
1. For $1 \leq i < j \leq n$, and for each increasing function $g$, if $c_i = c_j$, then $u_g^{(i)} < u_g^{(j)}$.
2. In our designations, $u_{QV}^{(1)} > u_{LV}^{(1)}$ and $u_{QV}^{(n)} < u_{LV}^{(n)}$.

3. If for some $1 \leq k \leq n$, $u_{QV}^{(k)} > u_{LV}^{(k)}$, then for each $1 \leq l < k$: $u_{QV}^{(l)} > u_{LV}^{(l)}$.

4. If for some $1 \leq k \leq n$, $u_{QV}^{(k)} < u_{LV}^{(k)}$, then for each $k < l \leq n$: $u_{QV}^{(l)} < u_{LV}^{(l)}$.

*In other words, the 1$^{st}$ statement says that VS (for any increasing g) preserves the increasing order of stake for voters, who vote for the same number of proposals. The 2$^{nd}$ says that for the voter with the largest stake, his RVR always decreases in QV-5, and for the voter with the smallest stake, his RVR always increases in QV-5, despite how many proposals they vote. The 3$^{rd}$ statement says that if for some voter his RVR in QV-5 is larger than for linear VS, then the same is true for all voters with smaller stake. Analogically, the 4$^{th}$ statement says that if for some voter his RVR in QV-5 is smaller than for linear VS, then the same is true for all voters with larger stake.*

*Proof.*

1. According to our designations, $s^{(i)} < s^{(j)}$, so $g(s^{(i)}) < g(s^{(j)})$ and $u_g^{(i)} = \frac{c_i \cdot g(s^{(i)})}{\sum_{j=1}^{n} c_j \cdot g(s^{(j)})} < \frac{c_i \cdot g(s^{(j)})}{\sum_{j=1}^{n} c_j \cdot g(s^{(j)})} = \frac{c_j \cdot g(s^{(j)})}{\sum_{j=1}^{n} c_j \cdot g(s^{(j)})} = u_g^{(j)}$.

2. According to our designations, $s^{(1)} < s^{(j)}$ for each $2 \leq j \leq n$. Then

$$u_{QV}^{(1)} = \frac{c_1 \cdot \sqrt{s^{(1)}}}{\sum_{j=1}^{n} c_j \cdot \sqrt{s^{(j)}}} = \frac{c_1}{\sum_{j=1}^{n} c_j \cdot \sqrt{\frac{s^{(j)}}{s^{(1)}}}} > \frac{c_1}{\sum_{j=1}^{n} c_j \cdot \frac{s^{(j)}}{s^{(1)}}} = \frac{c_1 \cdot s^{(1)}}{\sum_{j=1}^{n} c_j \cdot s^{(j)}} = u_{LV}^{(1)},$$

using that $\frac{s^{(j)}}{s^{(1)}} > 1$ and then $\frac{s^{(j)}}{s^{(1)}} > \sqrt{\frac{s^{(j)}}{s^{(1)}}}$.

Analogically, according to our designations, $s^{(n)} > s^{(j)}$ for each $1 \leq j \leq n-1$. Then

$$u_{QV}^{(n)} = \frac{c_n \cdot \sqrt{s^{(n)}}}{\sum_{j=1}^{n} c_j \cdot \sqrt{s^{(j)}}} = \frac{c_n}{\sum_{j=1}^{n} c_j \cdot \sqrt{\frac{s^{(j)}}{s^{(n)}}}} < \frac{c_n}{\sum_{j=1}^{n} c_j \cdot \frac{s^{(j)}}{s^{(n)}}} = \frac{c_n \cdot s^{(n)}}{\sum_{j=1}^{n} c_j \cdot s^{(j)}} = u_{LV}^{(n)},$$

using that $\frac{s^{(j)}}{s^{(n)}} < 1$ and then $\frac{s^{(j)}}{s^{(n)}} < \sqrt{\frac{s^{(j)}}{s^{(n)}}}$.

3. Let for some $1 \leq k \leq n$, $u_{QV}^{(k)} > u_{LV}^{(k)}$, which means that

$$u_{QV}^{(k)} = \frac{c_k \cdot \sqrt{s^{(k)}}}{\sum_{j=1}^{n} c_j \cdot \sqrt{s^{(j)}}} > u_{LV}^{(k)} = \frac{c_k \cdot s^{(k)}}{\sum_{j=1}^{n} c_j \cdot s^{(j)}}.$$

Then for each $1 \leq l < k$ we have:

$$u_{QV}^{(l)} = \frac{c_l \cdot \sqrt{s^{(l)}}}{\sum_{j=1}^{n} c_j \cdot \sqrt{s^{(j)}}} = \frac{\sqrt{s^{(l)}}}{\sqrt{s^{(k)}}} \cdot \frac{c_l}{c_k} \cdot \frac{c_k \cdot \sqrt{s^{(k)}}}{\sum_{j=1}^{n} c_j \cdot \sqrt{s^{(j)}}} > \sqrt{\frac{s^{(l)}}{s^{(k)}}} \cdot \frac{c_l}{c_k} \cdot u_{LV}^{(k)} >$$

$$> \frac{s^{(l)}}{s^{(k)}} \cdot \frac{c_l}{c_k} \cdot \frac{c_k \cdot s^{(k)}}{\sum_{j=1}^{n} c_j \cdot s^{(j)}} = \frac{c_l \cdot s^{(l)}}{\sum_{j=1}^{n} c_j \cdot s^{(j)}} = u_{LV}^{(l)},$$

using $\sqrt{\frac{s^{(l)}}{s^{(k)}}} > \frac{s^{(l)}}{s^{(k)}}$.

4. Let for some $1 \leq k \leq n$, $u_{QV}^{(k)} < u_{LV}^{(k)}$, which means that

$$u_{QV}^{(k)} = \frac{c_k \cdot \sqrt{s^{(k)}}}{\sum_{j=1}^{n} c_j \cdot \sqrt{s^{(j)}}} < u_{LV}^{(k)} = \frac{c_k \cdot s^{(k)}}{\sum_{j=1}^{n} c_j \cdot s^{(j)}}.$$

Then for each $k < l \leq n$ we have:

$$u_{QV}^{(l)} = \frac{c_l \cdot \sqrt{s^{(l)}}}{\sum_{j=1}^{n} c_j \cdot \sqrt{s^{(j)}}} = \frac{\sqrt{s^{(l)}}}{\sqrt{s^{(k)}}} \cdot \frac{c_l}{c_k} \cdot \frac{c_k \cdot \sqrt{s^{(k)}}}{\sum_{j=1}^{n} c_j \cdot \sqrt{s^{(j)}}} < \sqrt{\frac{s^{(l)}}{s^{(k)}}} \cdot \frac{c_l}{c_k} \cdot u_{LV}^{(k)} <$$

$$< \frac{s^{(l)}}{s^{(k)}} \cdot \frac{c_l}{c_k} \cdot \frac{c_k \cdot s^{(k)}}{\sum_{j=1}^{n} c_j \cdot s^{(j)}} = \frac{c_l \cdot s^{(l)}}{\sum_{j=1}^{n} c_j \cdot s^{(j)}} = u_{LV}^{(l)},$$

using $\sqrt{\frac{s^{(l)}}{s^{(k)}}} < \frac{s^{(l)}}{s^{(k)}}$.

Lemma is proved. □

**Generalization:** Proposition 2 is also true for arbitrary increasing function $g: R_{\geq 0} \to R_{\geq 0}$ with the next property: $g(x) \geq x$ for $x \in [0,1]$ and $g(x) \leq x$ for $x \geq 1$. For example, it is true for $g(x) = x^{\gamma}$ with arbitrary $\gamma \in (0,1)$.

### 3. Measuring decentralization in QV: Gini coefficient and Nakamoto index

As we proved above, when we change a linear voting system to a quadratic voting system, the largest stakeholders lose their relative voting power, but the smallest one's is increased. Intuition tells us, this means that the decentralization has increased. But to strictly prove the statement, we need to introduce the numerical characteristics of measures of decentralization.

There exists a significant number of such measures, described in the recent work by Mindaugas et al. (2024). Here we consider only two of them, which are the most widely used – Gini coefficient (GC) and Nakamoto coefficient (NC). Note that sometimes the word "index" is used instead of "coefficient". We are going to show that both of these characteristics become better when we move to Quadratic Voting from "ordinary" Linear Voting. Actually, we prove more general statements, which give tools to improve decentralization characteristics not only using square root of stakes, but arbitrary function with some definite properties.

#### 3.1. Gini coefficient for Quadratic voting system

One possible approach to measure the level of decentralization is to use the Gini coefficient. Below we define the Gini coefficient within the context of voting systems.

**Definition 3 (Gini coefficient).** Let $V = \{V^{(1)}, \ldots, V^{(n)}\}$ be the set of Voters, and $vc_g^{(1)} < \ldots < vc_g^{(n)}$ be their corresponding voting credits (with respect to function $g$), $vc_g^{(1)} + \ldots + vc_g^{(n)} = vc_g$, where $vc_g$ is their total voting credit.

Then the value of corresponding Gini coefficient is calculated as

$$G = G(s^{(1)}, \ldots, s^{(n)}) = \frac{2 \cdot \sum_{i=1}^{n} i \cdot vc_g^{(i)} - (n+1) \cdot vc_g}{n \cdot vc_g}. \quad (1)$$

The geometric sense of the Gini coefficient is explained in Fig. 2. It is equal to $\frac{A}{A+B}$, where $A$ is the value of square between Lorenz curve and line of equality, and $B$ is a square under the Line of equality.

The Lorenz line is the line which is built on the points with coordinates $(i, S_i)$, where $i \in \{1, \ldots, n\}$ and $S_i = \sum_{k=1}^{i} vc_g^{(k)}$. The line of equality is built on the points with coordinates $\left(i, \frac{i \cdot vc_g}{n}\right)$. These two lines coincides if and only if all voters have the same voting stake equal to $\frac{vc_g}{n}$. In this case $G = 0$. In the opposite case, when $vc_g^{(i)} = 0$ for $i = 1, \ldots, n-1$ and $vc_g^{(n)} = vc_g$, we have $G = 1$, which means that all voting power is concentrated in one voter.

**Lemma 1** (that geometric sense of Gini coefficient corresponds to Definition 3).
In our designations,

$$\frac{A}{A+B} = \frac{2 \cdot \sum_{i=1}^{n} i \cdot vc_g^{(i)} - (n+1) \cdot vc_g}{n \cdot vc_g}. \quad (2)$$

*Proof.* First of all, note that the Lorenz curve is smooth only in the continuous case. In the discrete case, like ours, it is a union of segments, that connects the points $(i, S_i)$ and $(i+1, S_{i+1})$, $i = \overline{1, n-1}$.

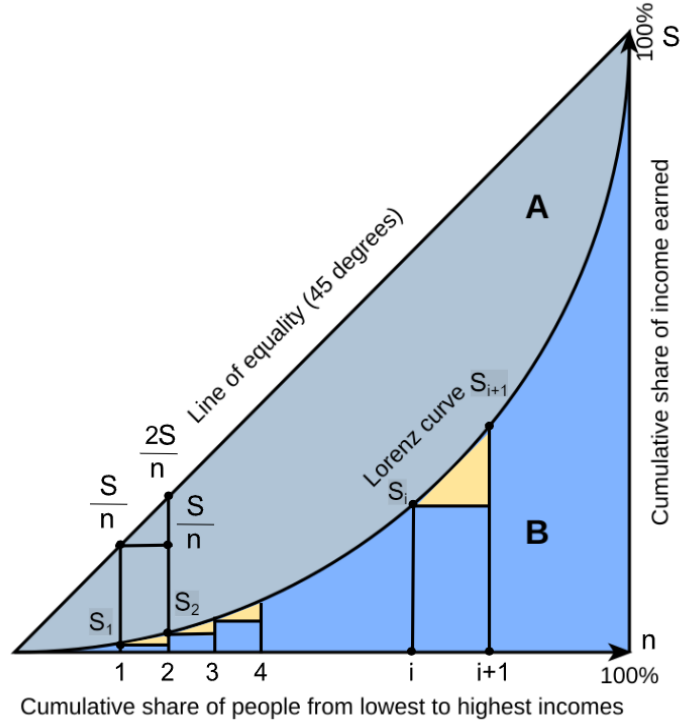

Fig. 2. Gini coefficient

Next, note that $A + B = \frac{vc_g \cdot n}{2}$, and to prove the lemma it's enough to calculate the area under the Lorenz curve, $S_L$. We divide these squares into parts, which are the trapeziums with vertices $(i, 0)$, $(i+1, 0)$, $(i, S_i)$, and $(i+1, S_{i+1})$, $i = \overline{1, n-1}$, and one triangle with vertices $(0,0)$, $(1,0)$, and $\left(1, vc_g^{(1)}\right)$. We divide $i$-th trapezium on rectangle with square $S_{i-1}$ and triangle with square $\frac{vc_g^{(i)}}{2}$. Then the sum of squares of these parts is equal to

$$S_L = \frac{vc_g^{(1)}}{2} + \sum_{i=2}^{n}\left(\sum_{k=1}^{i-1} vc_g^{(k)} + \frac{vc_g^{(i)}}{2}\right) = \frac{vc_g}{2} + \sum_{i=2}^{n}\left(\sum_{k=1}^{i-1} vc_g^{(k)}\right) = \frac{vc_g}{2} + + \sum_{i=1}^{n-1}(n-i) \cdot vc_g^{(i)}.$$

Then

$$A = (A + B) - B = \frac{n \cdot vc_g}{2} - \sum_{i=1}^{n-1}(n-i) \cdot vc_g^{(i)} - \frac{vc_g}{2} =$$

$$= \frac{(n-1) \cdot vc_g}{2} - n \cdot \sum_{i=1}^{n-1} vc_g^{(i)} + \sum_{i=1}^{n-1} i \cdot vc_g^{(i)} =$$

$$= \frac{(n-1) \cdot vc_g}{2} - n \cdot \left(vc_g - vc_g^{(n)}\right) + \sum_{i=1}^{n-1} i \cdot vc_g^{(i)}$$

and

$$\frac{A}{A+B} = \frac{2}{n \cdot vc_g} \cdot \left(\frac{(n-1) \cdot vc_g}{2} - n \cdot \left(vc_g - vc_g^{(n)}\right) + \sum_{i=1}^{n-1} i \cdot vc_g^{(i)}\right) =$$

$$= \frac{(n-1) \cdot vc_g - 2n \cdot \left(vc_g - vc_g^{(n)}\right) + 2 \cdot \sum_{i=1}^{n-1} i \cdot vc_g^{(i)}}{n \cdot vc_g} = \frac{2 \cdot \sum_{i=1}^{n-1} i \cdot vc_g^{(i)} - (n+1) \cdot vc_g}{n \cdot vc_g},$$

and the Lemma is proved.

Note that (2) may be rewritten as

$$\frac{A}{A+B} = G_{LV} = \frac{2 \cdot \sum_{i=1}^{n} i \cdot \frac{vc_g^{(i)}}{vc_g} - (n+1)}{n} = \frac{2 \cdot \sum_{i=1}^{n} i \cdot r_{LV}^{(i)} - (n+1)}{n}. \tag{3}$$

Also define $G_{QV}$ as

$$G_{QV} = \frac{2 \cdot \sum_{i=1}^{n} i \cdot r_{QV}^{(i)} - (n+1)}{n}. \tag{4}$$

The next Proposition shows that Quadratic Voting improves the decentralization properties of voting system in comparison with Linear Voting, if this property is measured using Gini coefficient.

**Proposition 3.** Let $V = \{V^{(1)}, \ldots, V^{(n)}\}$ be the set of Voters, and $s^{(1)} < \ldots < s^{(n)}$ be their corresponding voting stakes, $s^{(1)} + \ldots + s^{(n)} = s$. Then, in designations (3) and (4), the next inequality holds:
$$G_{QV} < G_{LV}.$$

*Proof:* due to (3) and (4), it`s enough to prove that
$$\sum_{i=1}^{n} i \cdot r_{LV}^{(i)} > \sum_{i=1}^{n} i \cdot r_{QV}^{(i)}. \tag{5}$$

To continue the proof of Proposition 3, we need to formulate the next Lemma.

**Lemma 2.** Let $0 \leq y_1 \leq y_2 \leq \ldots \leq y_n$, $0 \leq a_1 \leq a_2 \leq \ldots \leq a_n$, and $0 \leq b_1 \leq b_2 \leq \ldots \leq b_n$ are such that their partial sums $A_k = \sum_{i=1}^{k} a_i$ and $B_k = \sum_{i=1}^{k} b_i$ satisfy the conditions:
(i) $A_k \leq B_k$ for $1 \leq k < n$;
(ii) $A_n = B_n$.
Then $\sum_{i=1}^{n} a_i y_i \geq \sum_{i=1}^{n} b_i y_i$.

*Proof of Lemma 2:* using summation by parts, we obtain:
$$\sum_{i=1}^{n} a_i y_i = \sum_{i=1}^{n-1} A_i (y_i - y_{i+1}) + A_n y_n \geq \sum_{i=1}^{n-1} B_i (y_i - y_{i+1}) + B_n y_n = \sum_{i=1}^{n} b_i y_i,$$
where we use that $y_i - y_{i+1} \leq 0$, and the Lemma is proved.

*Proof of Proposition 3* (continuation).

Below we prove a more general statement.

Let $\gamma \in (0,1)$, and set $a_k = \frac{s^{(k)}}{\sum_{i=1}^{n} s^{(i)}}$, $b_k = \frac{(s^{(k)})^\gamma}{(\sum_{i=1}^{n} s^{(i)})^\gamma}$. Now we are going to show that the partial sums $A_k = \sum_{i=1}^{k} a_i$ and $B_k = \sum_{i=1}^{k} b_i$ satisfy the conditions of Lemma 2. It`s easy to see that $A_n = B_n = 1$. To prove that $A_k \leq B_k$ for $1 \leq k < n$ we prove the equivalent inequality that $\frac{1}{A_k} > \frac{1}{B_k}$:

$$\frac{1}{A_k} = \frac{\sum_{i=1}^{n} s^{(i)}}{\sum_{i=1}^{k} s^{(i)}} = 1 + \frac{\sum_{i=k+1}^{n} s^{(i)}}{\sum_{i=1}^{k} s^{(i)}} = 1 + \frac{\sum_{i=k+1}^{n} \frac{s^{(i)}}{s^{(k+1)}}}{\sum_{i=1}^{k} \frac{s^{(i)}}{s^{(k+1)}}} > 1 + \frac{\sum_{i=k+1}^{n} \left(\frac{s^{(i)}}{s^{(k+1)}}\right)^\gamma}{\sum_{i=1}^{k} \left(\frac{s^{(i)}}{s^{(k+1)}}\right)^\gamma} = \frac{1}{B_k},$$

using that for $i \geq k+1$ we have $\frac{s^{(i)}}{s^{(k+1)}} \geq 1$ and for $i \leq k$ we have $\frac{s^{(i)}}{s^{(k+1)}} < 1$.

Then, setting in the Lemma 2 $y_i = i$, we have
$$\sum_{i=1}^{n} i \cdot r_{LV}^{(i)} > \sum_{i=1}^{n} i \cdot \frac{(r_{LV}^{(i)})^\gamma}{\sum_{l=1}^{n} (r_{LV}^{(l)})^\gamma}.$$

In particular case when $\gamma = \frac{1}{2}$ we have the inequality (5), and Proposition is proved.

From the proof on Proposition 3, we can obtain a more general statement.

**Proposition 4.** For $\gamma \in (0,1)$, define
$$r_{\gamma V}^{(i)} = \frac{(s^{(i)})^\gamma}{\sum_{i=1}^{n} (s^{(i)})^\gamma} \text{ and } G_{\gamma V} = \frac{2 \cdot \sum_{i=1}^{n} i \cdot r_{\gamma V}^{(i)} - (n+1)}{n}.$$
Then $G_{\gamma V} < G_{LV}$.

The Proposition 4 states that Gini coefficient, defined for any power $\gamma \in (0,1)$ of corresponding stares, is smaller than "linear" Gini coefficient. It means that applying arbitrary power $\gamma \in (0,1)$ (not only $\gamma = \frac{1}{2}$) improves decentralization properties of VS.

**Note** that for QV-1 the result about decentralization increasing, generally speaking, isn`t true. More precisely, it depends on the stake distribution, done by each voter. For example, if a Voter with stake $m$ units gives all his votes to one proposal, his voting ratio is $\sqrt{m}$, as for QV-2. But if he gives 1 vote for each of $m$ proposals, his voting ratio is $m$, as in a linear voting system.

In many practical applications, QV encourages deeper collaboration among participants, fostering the formation of mutually beneficial agreements prior to the voting process. However, the observation above illustrates that large stakeholders ("whales") can coordinate strategically to maintain disproportionate influence - thereby undermining decentralization. We refer to this behavior as a collusion attack. We describe it using the following example. There are 3 proposals, and Voter 1 is interested only in proposal number 1, Voter 2 is interested only in proposal number 2, and Voter 3 is interested only in proposal number 3. Let each of them have 3 coins. Then, if each voter voted for the proposal he is interested in, his impact on voting for this proposal would be $\sqrt{3}$. But if they agreed to give 1 vote for each of proposal numbers 1, 2, and 3, then each of these proposals got 3 votes.

### 3.2. Nakamoto coefficient for Quadratic voting system

Informally, the Nakamoto coefficient may be defined as the minimum number of entities in some system needed to control the decison-making of the system. As a threshold, the value of 51% of the total capacity is often (mistakenly) used (Srinivasan and Lee 2017), especially when talking about blockchain. In what follows, we need to formalize this notion, to create strict proofs of its properties. We will give two definitions of the Nakamoto coefficient – as a number of entities (classical case) and as the ratio of the total number of entities (alternative case), where the former definition is more usual, but the latter is more informative.

**Definition 4 (Nakamoto coefficient).** Let $V = \{V^{(1)}, \ldots, V^{(n)}\}$ be the set of Voters, and $vc_g^{(1)} < \ldots < vc_g^{(n)}$ be their corresponding voting credits (with respect to function $g$), $vc_g^{(1)} + \ldots + vc_g^{(n)} = vc_g$, where $vc_g$ is their total voting credit. Let $a \in 0,1$ be such threshold value that the stake amount $as$ has full control over the system.

Then the value of the Nakamoto coefficient, corresponding to the threshold $a$, is calculated as
$$N^{(a)} = N^{(a)}\big(vc_g^{(1)}, \ldots, vc_g^{(n)}\big) = min\{k: \sum_{n-k+1}^{n} vc_g^{(i)} \geq a \cdot vc_g\} \quad (6)$$
for "classical" case, or as
$$\widetilde{N}^{(a)} = \frac{1}{n} N^{(a)}\big(vc_g^{(1)}, \ldots, vc_g^{(n)}\big) = min\left\{\frac{k}{n}: \sum_{n-k+1}^{n} \frac{vc_g^{(i)}}{vc_g} \geq a\right\} \quad (7)$$
in an alternative case.

Note that the larger the Nakamoto coefficient, the more decentralization is achieved in the system. The NC, defined according to (6), takes (integer) values from 1 to $n$; if it is defined according to (7), it takes values from in the interval (0,1). It is easy to see that if (6) increases, then so does (7). So, it does not matter what definition is used. In this subsection, we will use (6).

The next Proposition shows that, like in the case of the Gini coefficient, the decentralization property for a Quadratic voting system is better than for a linear voting system, if this property is measured using the Nakamoto coefficient.

If we move from a linear voting system to a quadratic voting system and apply the square root function to stakes, then the Nakamoto coefficient for a Quadratic voting system may be calculated as
$$N_{QV}^{(a)} = N^{(a)}\left(\sqrt{s^{(1)}}, \ldots, \sqrt{s^{(n)}}\right) = min\left\{k: \sum_{i=n-k+1}^{n} \sqrt{s^{(i)}} \geq a \sum_{i=1}^{n} \sqrt{s^{(i)}}\right\} \quad (8)$$
for "classical" case, or as
$$\widetilde{N}_{QV}^{(a)} = \frac{1}{n} N^{(a)}\left(\sqrt{s^{(1)}}, \ldots, \sqrt{s^{(n)}}\right) = min\left\{\frac{k}{n}: \sum_{i=n-k+1}^{n} \frac{\sqrt{s^{(i)}}}{\sum_{i=1}^{n} \sqrt{s^{(i)}}} \geq a\right\} \quad (9)$$
in an alternative case.

If in the formula (6) $g(x) = x$, we will use designation $N_{LV}^{(a)}$, and if $g(x) = x^\gamma$, where $\gamma \in (0,1)$, we will write
$$N_{\gamma V}^{(a)} = N^{(a)}\big((s^{(1)})^\gamma, \ldots, (s^{(n)})^\gamma\big) = min\left\{k: \sum_{i=n-k+1}^{n} \frac{(s^{(i)})^\gamma}{\sum_{i=1}^{n} (s^{(i)})^\gamma} \geq a\right\}. \quad (10)$$

**Proposition 5.** Let $V = \{V^{(1)}, \ldots, V^{(n)}\}$ be the set of Voters, and $s^{(1)} < \ldots < s^{(n)}$ be their corresponding voting stakes, $s^{(1)} + \ldots + s^{(n)} = s$. Then, in designations (6) and (7), for arbitrary $a \in (0,1)$ the next inequality holds:

$$N_{\gamma V}{}^{(a)} > N_{LV}{}^{(a)}.$$

*Proof:* it`s enough to prove that for arbitrary $k \in [n]$:

$$\sum_{i=n-k+1}^{n} \frac{(s^{(i)})^\gamma}{\sum_{i=1}^{n}(s^{(i)})^\gamma} < \sum_{i=n-k+1}^{n} \frac{s^{(i)}}{\sum_{i=1}^{n} s^{(i)}}. \tag{11}$$

As in proof of Proposition 3, define $a_k = \frac{s^{(k)}}{\sum_{i=1}^{n} s^{(i)}}$, $b_k = \frac{(s^{(k)})^\gamma}{(\sum_{i=1}^{n} s^{(i)})^\gamma}$. It was shown in the proof, the partial sums $A_k = \sum_{i=1}^{k} a_i$ and $B_k = \sum_{i=1}^{k} b_i$ satisfy the conditions of Lemma 2. Then, setting in the Lemma 2

$$y_i = \begin{cases} 0, for\ i \in [n-k] \\ 1, for\ i \in \{n-k+1, \ldots, n\} \end{cases}$$

obtain $\sum_{i=n-k+1}^{n} a_i \geq \sum_{i=n-k+1}^{n} b_i$, or $\sum_{i=n-k+1}^{n} \frac{(s^{(i)})^\gamma}{\sum_{i=1}^{n}(s^{(i)})^\gamma} < \sum_{i=n-k+1}^{n} \frac{s^{(i)}}{\sum_{i=1}^{n} s^{(i)}}$.

The Lemma is proven.

### 4. Utility maximization

In this section, we consider QV-1 and QV-2 and show that it is very important to keep voting results secret till the moment when all voters make their choices. In the opposite case, the voter who votes after all other voters can maximize his utility, by analyzing the results of voting he can see.

Below we generalize results of Dimitri (2022) about maximization of the utility function corresponding to a voter.

#### 4.1. Utility maximization for QV-1

Let's consider QV-1, where $g(x) = x$, so $vc_g{}^{(i)} = s^{(i)}$.

We consider the set of voters $V = \{V^{(1)}, \ldots, V^{(n)}\}$ with stakes $s^{(1)}, \ldots, s^{(n)}$, where $s^{(1)} + \ldots + s^{(n)} = s$. Let $m$ be the number of proposals, and voter $V^{(i)}$ divides his stake $s^{(i)}$ between these $m$ proposals as $(s_1{}^{(i)}, \ldots, s_m{}^{(i)})$ in a such way, that $s^{(i)} = (s_1{}^{(i)})^2 + \ldots + (s_m{}^{(i)})^2$ (we may consider both positive and negative values of $s_r{}^{(i)}$, or only non-negative values).

Following the mentioned article (Dimitri, 2022), we define $\pi_r{}^{(i)}$ the profit of $V^{(i)}$ which he can obtain when voting for proposal with number $r$. Next, for each $1 \leq r \leq m$ define the probability $P_r{}^{(i)}$ for $V^{(i)}$ to get the profit $\pi_r{}^{(i)}$ as

$$P_r{}^{(i)} = \frac{s_r{}^{(i)} + a_r}{s_r{}^{(i)} + b_r}, \tag{12}$$

where $a_r$ is the value of total stake (except $s_r{}^{(i)}$) which votes for this proposal in the same as $s_r{}^{(i)}$, and $b_r$ is the value of total stake (except $s_r{}^{(i)}$) which votes for this proposal. Note that in (12) we generalize the assumption made by Posner and Weyl (2014), that all other voters vote other than $V^{(i)}$. Then the utility function $U^{(i)} = U^{(i)}(s_1{}^{(i)}, \ldots, s_m{}^{(i)})$ for $V^{(i)}$ may be written as

$$U^{(i)}(s_1{}^{(i)}, \ldots, s_m{}^{(i)}) = \sum_{i=1}^{m} \left\{ v_r{}^{(i)} \cdot \frac{s_r{}^{(i)} + a_r}{s_r{}^{(i)} + b_r} \right\}. \tag{13}$$

Before proving the proposition about utility maximization, we need to prove auxiliary Lemma.

**Lemma 3.** Let quadratic form $A = (a_{ij})_{i,j=1}^{n+1}$ is as follows:

1) $a_{ij} = a_{ji}, i, j = \overline{1, n+1}$;
2) $a_{ij} = 0 \Leftrightarrow i \neq j$, for $i, j = \overline{1, n}$ or $i = j = n+1$;
3) for all other $i, j$: $a_{ij} < 0$.

Then $A = (a_{ij})_{i,j=1}^{n+1}$ is negatively defined.

*Proof.* It is more convenient to write $A = (a_{ij})_{i,j=1}^{n+1}$ in the form with elements $-a_1, -a_2, \ldots, -a_n, 0$ on the main diagonal, with elements $-b_1, -b_2, \ldots, -b_n, 0$ in the last $n+1$-th row and with the same elements $-b_1, -b_2, \ldots, -b_n, 0$ in the last $n + 1$-th column. Also define $A_k = (a_{ij})_{i,j=1}^k$ the sequence of its submatrixes. Let $D_k = det\, A_k = d_k \prod_{i=1}^k (-a_i)$. Then:

$$D_{k+1} = -a_{k+1} D_k - (-1)^k \cdot b_{k+1}^2 \cdot \prod_{i=1}^k a_i; \qquad d_{k+1} = d_k + \frac{b_{k+1}^2}{a_{k+1}}; \qquad d_k = \sum_{i=1}^k \frac{b_i^2}{a_i},$$

so the quadratic form is negatively defined.

### Proposition 5 (Utility maximization for QV-1).

Let (13) be a continuous function. Then, if there exists a set of $s_r^{(i)}$, $1 \leq r \leq m$, such that the solution of the system of equations

$$\pi_r^{(i)} \cdot \frac{b_r - a_r}{(s_r^{(i)} + b_r)^2} - \sqrt{\frac{\sum_{r=1}^m \left\{ \pi_r^{(i)} \cdot \frac{b_r - a_r}{(s_r^{(i)} + b_r)^2} \right\}^2}{s^{(i)}}} s_r^{(i)} = 0, \, 1 \leq r \leq m, \qquad (14)$$

then this solution maximizes the utility function (13) under the condition $s^{(i)} = (s_1^{(i)})^2 + \ldots + (s_m^{(i)})^2$.

In the case when the solution of (14) does not exist, the maximum of function (13) still exists, since the maximum of a continuous function is a closed area.

In other words, voter $V^{(i)}$, who knows how the stake of other voters is distributed among proposals (i.e. knows $(a_r, b_r)$ for $1 \leq r \leq m$), can maximize his utility function (13).

*Proof.* We need to find the maximum of the function (13) under the condition $s^{(i)} = (s_1^{(i)})^2 + \ldots + (s_m^{(i)})^2$. The corresponding Lagrange function is

$$L^{(i)}(s_1^{(i)}, \ldots, s_m^{(i)}, \lambda^{(i)}) = \sum_{i=1}^m \pi_r^{(i)} \cdot \frac{s_r^{(i)} + a_r}{s_r^{(i)} + b_r} - \lambda^{(i)} \left( \sum_{r=1}^m (s_r^{(i)})^2 - s^{(i)} \right). \qquad (15)$$

The partial derivations are

$$\frac{\partial L^{(i)}(s_1^{(i)}, \ldots, s_m^{(i)}, \lambda^{(i)})}{\partial s_r^{(i)}} = \pi_r^{(i)} \cdot \frac{b_r - a_r}{(s_r^{(i)} + b_r)^2} - 2\lambda^{(i)} s_r^{(i)} = 0, 1 \leq r \leq m;$$

$$\frac{\partial L^{(i)}(s_1^{(i)}, \ldots, s_m^{(i)}, \lambda^{(i)})}{\partial \lambda^{(i)}} = \sum_{r=1}^m (s_r^{(i)})^2 - s^{(i)} = 0,$$

from where we get

$$\lambda^{(i)} = \sqrt{\frac{\sum_{r=1}^m \left\{ \pi_r^{(i)} \cdot \frac{b_r - a_r}{2(s_r^{(i)} + b_r)^2} \right\}^2}{s^{(i)}}}$$

and $s_r^{(i)}$, $1 \leq r \leq m$, are the solution of the equation

$$\pi_r^{(i)} \cdot \frac{b_r - a_r}{(s_r^{(i)} + b_r)^2} - 2 \sqrt{\frac{\sum_{r=1}^m \left\{ \pi_r^{(i)} \cdot \frac{b_r - a_r}{2(s_r^{(i)} + b_r)^2} \right\}^2}{s^{(i)}}} s_r^{(i)} = 0.$$

Assume that this solution exists and prove that it maximizes the function (13). Find the second derivative as a quadratic form and prove that it is negatively defined. Note that

$$\frac{\partial^2 L^{(i)}(s_1^{(i)}, \ldots, s_m^{(i)}, \lambda^{(i)})}{\partial s_r^{(i)} \partial s_k^{(i)}} = \begin{cases} 0, \text{if } r \neq k; \\ -2\pi_r^{(i)} \cdot \frac{b_r - a_r}{(s_r^{(i)} + b_r)^3} - 2\lambda^{(i)} < 0, r = k, \end{cases}$$

$$\frac{\partial^2 L^{(i)}(s_1^{(i)}, \ldots, s_m^{(i)}, \lambda^{(i)})}{\partial s_r^{(i)} \partial \lambda^{(i)}} = -2s_r^{(i)} < 0.$$

So, the corresponding quadratic form satisfies the condition of Lemma 3 and is negatively defined. The Proposition is proved.

### 4.2. Utility maximization for QV-2

Below we use the same notations that in 4.1 with only difference – condition

$$vc_{QV}^{(i)} = \sqrt{s^{(i)}} = s_1^{(i)} + \ldots + s_m^{(i)}$$

instead of $s^{(i)} = \left(s_1^{(i)}\right)^2 + \ldots + \left(s_m^{(i)}\right)^2$.

**Proposition 6 (Utility maximization for QV-2).**

Let (13) be a continuous function. Then, if there exists such set of $s_r^{(i)}$, $1 \leq r \leq m$, which is the solution of the equations system

$$\pi_r^{(i)} \cdot \frac{b_r - a_r}{(s_r^{(i)} + b_r)^2} - \frac{1}{\sqrt{s^{(i)}}} \cdot \sum_{r=1}^m \pi_r^{(i)} \cdot \frac{s_r^{(i)} \cdot (b_r - a_r)}{(s_r^{(i)} + b_r)^2} = 0, \; 1 \leq r \leq m, \quad (16)$$

this solution maximizes the utility function (13) under the condition $\sqrt{s^{(i)}} = s_1^{(i)} + \ldots + s_m^{(i)}$.

In case when the solution of (16) does not exist, the maximum of function (13) still exists, as the maximum of continuous function in a closed area.

In other words, voter $V^{(i)}$, who knows how stake of other voters is distributed among proposals (i.e. knows $(a_r, b_r)$ for $1 \leq r \leq m$), can maximize his utility function (13).

*Proof.* We need to find the maximum of the function (13) under the condition $s^{(i)} = \left(s_1^{(i)}\right)^2 + \ldots + \left(s_m^{(i)}\right)^2$. The corresponding Lagrange function is

$$L^{(i)}\left(s_1^{(i)}, \ldots, s_m^{(i)}, \lambda^{(i)}\right) = \sum_{i=1}^m \pi_r^{(i)} \cdot \frac{s_r^{(i)} + a_r}{s_r^{(i)} + b_r} - \lambda^{(i)} \left(\sum_{r=1}^m s_r^{(i)} - \sqrt{s^{(i)}}\right). \quad (17)$$

The partial derivations are

$$\frac{\partial L^{(i)}\left(s_1^{(i)}, \ldots, s_m^{(i)}, \lambda^{(i)}\right)}{\partial s_r^{(i)}} = \pi_r^{(i)} \cdot \frac{b_r - a_r}{(s_r^{(i)} + b_r)^2} - 2\lambda^{(i)} = 0, \; 1 \leq r \leq m;$$

$$\frac{\partial L^{(i)}\left(s_1^{(i)}, \ldots, s_m^{(i)}, \lambda^{(i)}\right)}{\partial \lambda^{(i)}} = \sum_{r=1}^m s_r^{(i)} - \sqrt{s^{(i)}} = 0,$$

from where we get

$$\lambda^{(i)} = \frac{1}{\sqrt{s^{(i)}}} \cdot \sum_{r=1}^m \pi_r^{(i)} \cdot \frac{s_r^{(i)} \cdot (b_r - a_r)}{2(s_r^{(i)} + b_r)^2}$$

and $s_r^{(i)}$, $1 \leq r \leq m$, are the solution of the equations system

$$\pi_r^{(i)} \cdot \frac{b_r - a_r}{(s_r^{(i)} + b_r)^2} - 2 \frac{1}{\sqrt{s^{(i)}}} \cdot \sum_{r=1}^m \pi_r^{(i)} \cdot \frac{s_r^{(i)} \cdot (b_r - a_r)}{2(s_r^{(i)} + b_r)^2} = 0, \; 1 \leq r \leq m.$$

Assume that this solution exists and prove that it maximizes the function (13). Find the second derivative as a quadratic form and prove that it is negatively defined. Note that

$$\frac{\partial^2 L^{(i)}\left(s_1^{(i)}, \ldots, s_m^{(i)}, \lambda^{(i)}\right)}{\partial s_r^{(i)} \partial s_k^{(i)}} = \begin{cases} 0, \text{ if } r \neq k; \\ -2\pi_r^{(i)} \cdot \frac{b_r - a_r}{(s_r^{(i)} + b_r)^3} < 0, r = k, \end{cases}$$

$$\frac{\partial^2 L^{(i)}\left(s_1^{(i)}, \ldots, s_m^{(i)}, \lambda^{(i)}\right)}{\partial s_r^{(i)} \lambda^{(i)}} = -2 < 0.$$

So, the corresponding quadratic form satisfies the condition of Lemma 3 and is negatively defined. The Proposition is proved.

## 5. Reducing the initial stake impact of "whales" to the desired level

As it was introduced above, we consider the set of voters $V = \{V^{(1)}, \ldots, V^{(n)}\}$ with stakes $s^{(1)}, \ldots, s^{(n)}$, where $s^{(1)} < \ldots < s^{(n)}$ and $s^{(1)} + \ldots + s^{(n)} = s$.

For some non-decreasing function $T: R_+ \to R_+$, define corresponding transformation $T$ of initial stake distribution as

$$T: \{s^{(1)}, \ldots, s^{(n)}\} \to \{\tilde{s}^{(1)}, \ldots, \tilde{s}^{(n)}\}, \text{ with } T(s^{(i)}) = \tilde{s}^{(i)}, i \in [n]. \quad (18)$$

**Definition 5.** We define *the initial relative stake impact of Voter $V^{(k)}$* using the stake ratio $s_{rel}^{(k)} = \frac{s^{(k)}}{\sum_{i=1}^n s^{(i)}}$, and *the relative T-transformed stake impact of Voter $V^{(k)}$* as $s_{rel,T}^{(k)} = \frac{T(s^{(k)})}{\sum_{i=1}^n T(s^{(i)})}$.

Without being tied to a specific type of voting, consider the next problem: we want to create a transformation $T$ of the initial stake distribution, which has the following properties:

1) if $s^{(i)} < s^{(j)}$, then $s_{rel,T}{}^{(i)} < s_{rel,T}{}^{(j)}$, i.e. if the initial stake of $V^{(i)}$ was smaller than the initial stake of $V^{(j)}$, then the same inequality holds for their relative $T$-transformed stake impacts;

2) $s_{rel}{}^{(1)} < s_{rel,T}{}^{(1)}$ and $s_{rel}{}^{(n)} < s_{rel,T}{}^{(n)}$, i.e. the Voter with the smallest stake increases his power after the transformation, and the Voter with the largest stake decreases his power after the transformation;

3) if for some $k \in [n]$ we have $s_{rel}{}^{(k)} < s_{rel,T}{}^{(k)}$, then for each $l \in [k]$: $s_{rel}{}^{(l)} < s_{rel,T}{}^{(l)}$, i.e. if voter $V^{(k)}$ increases his relative stake impact after the transformation, than all voters with smaller stakes also increase their relative stake impacts after the transformation;

4) if for some $k \in [n]$ we have $s_{rel}{}^{(k)} > s_{rel,T}{}^{(k)}$, then for each $l \in [k, n]$: $s_{rel}{}^{(l)} > s_{rel,T}{}^{(l)}$, i.e. if voter $V^{(k)}$ decreases his relative stake impact after the transformation, than all voters with larger stakes also decrease their relative stake impacts after the transformation;

5) for stake distribution $\{\tilde{s}^{(1)}, \ldots, \tilde{s}^{(n)}\}$, obtained after transformation, Gini index and Nakamoto index show better decentralization;

6) for given value $\alpha \in (0, s_{rel}{}^{(n)})$ the next inequality holds:
$$s_{rel,T}{}^{(i)} \leq \alpha, i \in [n].$$

From Propositions 1-5, and generalization of Proposals 1 and 2, we obtain the family of transformations, indexed with $\gamma \in (0,1)$ and satisfied properties 1-5:
$$T_\gamma(s^{(i)}) = (s^{(i)})^\gamma. \tag{19}$$
For arbitrary fixed stake distribution $S = \{s^{(1)}, \ldots, s^{(n)}\}$, define function $t_S(\gamma): [0,1] \to R$ as
$$t_S(\gamma) = \frac{(s^{(n)})^\gamma}{\sum_{i=1}^n (s^{(i)})^\gamma}, \tag{20}$$
equal to the largest relative $T_\gamma$-transformed stake impact.

The next Proposition shows that the transformation (19) also satisfies the property 6.

**Proposition 7.** In our designations, for arbitrary given $\alpha \in \left(\frac{1}{n}, s_{rel}{}^{(n)}\right)$, there exists unique $\gamma = \gamma(\alpha) \in (0,1)$ such that for corresponding transformation $T_\gamma$, where $T_\gamma(s^{(i)}) = (s^{(i)})^\gamma$, $i \in [n]$, the next conditions hold:
$$s_{rel,T_\gamma}{}^{(n)} = \alpha \text{ and } s_{rel,T_\gamma}{}^{(k)} < \alpha \text{ for } j \in [n-1].$$

*Proof.* Note that the function $t_S(\gamma)$, $\gamma \in [0,1]$, defined in (20), is continuously differentiable, as composition of continuously differentiable functions.

It's easy to prove that function (20) is increasing. Indeed,
$$t_S'(\gamma) = \frac{(s^{(n)})^\gamma \cdot \ln s^{(n)} \cdot \sum_{i=1}^n (s^{(i)})^\gamma - (s^{(n)})^\gamma \cdot \sum_{i=1}^n \{(s^{(i)})^\gamma \cdot \ln s^{(i)}\}}{\left(\sum_{i=1}^n (s^{(i)})^\gamma\right)^2},$$
and it's enough to prove that
$$(s^{(n)})^\gamma \cdot \ln s^{(n)} \cdot \sum_{i=1}^n (s^{(i)})^\gamma > (s^{(n)})^\gamma \cdot \sum_{i=1}^n \{(s^{(i)})^\gamma \cdot \ln s^{(i)}\},$$
which is the same as
$$\ln s^{(n)} \cdot \sum_{i=1}^n (s^{(i)})^\gamma > \sum_{i=1}^n \{(s^{(i)})^\gamma \cdot \ln s^{(i)}\}. \tag{21}$$
As $s^{(i)} \leq s^{(n)}$, $i \in [n]$, then $\ln s^{(i)} \leq \ln s^{(n)}$, $i \in [n]$, so (21) holds, and function (20) is increasing.

This fact means that we have the next property of $T_\gamma$-transformation:

- the smaller is $\gamma$, the smaller is the largest relative $T_\gamma$-transformed stake impact.

Next, $t_S(0) = \frac{1}{n}$ and $t_S(1) = s_{rel}{}^{(n)}$. So, by Lagrange theorem, when $\gamma$ changes from 0 to 1, the function (20) takes all values from $\frac{1}{n}$ to $s_{rel}{}^{(n)}$. Then for arbitrary $\alpha \in \left(\frac{1}{n}, s_{rel}{}^{(n)}\right)$ there exists $\gamma = \gamma(\alpha) \in (0,1)$ such that $t_S(\gamma) = \alpha$, which means that $s_{rel,T_\gamma}{}^{(n)} = \alpha$, and such $\gamma$ is unique because of monotonicity of function $t_S(\gamma)$. As for all $i \in [n]$ we have $s_{rel,T_\gamma}{}^{(i)} \leq s_{rel,T_\gamma}{}^{(n)}$, then $s_{rel,T_\gamma}{}^{(i)} \leq \alpha$, $i \in [n]$, and the proposition is proved. □

**Note:** We may generalize Proposition 7 for the case of several largest stakeholders, considering the function

$$t_{S,k}(\gamma) = \frac{\sum_{i=k}^{n}(s^{(i)})^{\gamma}}{\sum_{i=1}^{n}(s^{(i)})^{\gamma}} \qquad (22)$$

instead of (20). Here $k$ is the number of the largest stakeholders, which total stake impact we need to restrict.

For arbitrary given stake distribution, Proposition 7 and its generalization give us instruments to achieve the desired level of maximal stake impact, using stake transformation with properties 1)-6). We may use the next algorithm for this purpose.

**Algorithm for achieving desired level of maximal stake impact.**
**Input:** stake distribution $S = \{s^{(1)}, \ldots, s^{(n)}\}$ (where $s^{(1)} < \ldots < s^{(n)}$);
number $k$ of the largest stakeholders;
desired level $\alpha \in (0,1)$.
**Step 1.** Calculate $s_{rel,k}^{(n)} = \frac{\sum_{i=k}^{n} s^{(i)}}{\sum_{i=1}^{n} s^{(i)}}$.
**Step 2.** If $\alpha > s_{rel,k}^{(n)}$ then print "incorrect input"; stop.
**Step 3.** Using dichotomy method (or other method), find such $\gamma \in (0,1)$ that

$$t_{S,k}(\gamma) = \frac{\sum_{i=k}^{n}(s^{(i)})^{\gamma}}{\sum_{i=1}^{n}(s^{(i)})^{\gamma}} \approx \alpha,$$

with appropriate approximation in a given range (set initially $\gamma = 1$ and then decrease it).
**Output:** a new transformed stake distribution $\{\tilde{s}^{(1)}, \ldots, \tilde{s}^{(n)}\}$, with $\tilde{s}^{(i)} = T_{\gamma}(s^{(i)}) = (s^{(i)})^{\gamma}$, $i \in [n]$.

### Discussion and Conclusions

This research explores the dynamics of Quadratic Voting and its generalization, focusing on the impact on decentralization and fairness in blockchain governance systems. We provide significant theoretical and practical insights into the mechanisms of QV and its adaptability, addressing key challenges associated with traditional linear voting schemes. We mainly focus on Quadratic Voting with a split stake (with Yes-Abstain) option as the most common and popular alternative scheme.

The presented formal analysis demonstrates that QV – particularly Types 2 and 3 – and its generalization enhance decentralization as confirmed by both the Gini coefficient and the Nakamoto coefficient metrics. These findings establish QV as a viable alternative to linear voting by better balancing voting ratio distribution. A notable discovery is the identification of a "threshold" stakeholder under QV, where participants with stakes below this threshold experience an increase in their relative voting ratio, while those above it see a reduction. This threshold can be flexibly adjusted in the generalized QV model through parameter tuning, enabling the design of systems where, for example, 40% of the wealthiest stakeholders hold no more than 60% of the votes.

Importantly, fairness remains intact under QV. While stakeholders with higher stakes continue to hold a proportionally greater voting ratio, the enhanced decentralization ensures a more equitable distribution of influence compared to linear voting. These properties make QV particularly appealing for decentralized systems aiming to empower smaller participants without undermining the fundamental principle that greater stakes merit greater influence.

While QV and its generalizations offer significant advantages, their deployment in decentralized systems requires careful implementation with privacy-preserving cryptographic voting protocols. Without these safeguards, fairness cannot be guaranteed; for instance, voters casting their ballots after observing earlier votes could manipulate outcomes, undermining the integrity of the voting process. This highlights the importance of integrating robust cryptographic solutions to ensure transparency, fairness, and resistance to strategic exploitation.

The research also demonstrates that large stakeholders can strategically coordinate to retain disproportionate influence in Type 1 QV – attempting collusion-based attacks. This highlights the need for robust resilience mechanisms, as emphasized in our paper.

Beyond theoretical proofs, this study introduces several novel contributions:

1. A formal model for the Gini coefficient and Nakamoto coefficient, with proofs validating their applicability to QV.
2. Evidence that Type 2 QV achieves greater equality for voters compared to the "1 coin – 1 vote" scheme under specific conditions.
3. Discovery of a strict split among voters based on their relative voting ratio in QV.
4. Generalization of QV, allowing the use of fractional roots (e.g., any power from 0 to 1) to enhance voter equality.
5. Proposal of an algorithm to achieve a desired level of maximum stake impact, addressing concerns around concentration of influence.

**Practical application**. The generalized QV model introduced in this research enables algorithmic parametrization to achieve tailored levels of decentralization based on specific use cases. This flexibility makes it applicable across diverse domains, including:
- enhancing user interaction with cryptocurrency platforms;
- facilitating community events and educational initiatives;
- supporting charitable activities through decentralized decision-making.

This study establishes Quadratic Voting and its generalizations as powerful tools for promoting decentralization and fairness in blockchain governance. By addressing vulnerabilities, enhancing resilience, and introducing a flexible framework for parameter adjustment, our findings provide a robust foundation for the adoption of QV in decentralized systems. Future work will focus on refining these mechanisms, exploring additional applications, and expanding the theoretical understanding of their properties to further support the evolution of decentralized governance models.